\RequirePackage{fix-cm}
\documentclass[journal]{IEEEtran}                  
\usepackage{amssymb,amsfonts}
\usepackage{algorithmic}
\usepackage{graphicx}
\usepackage{subcaption}
\usepackage{textcomp}
\usepackage[dvipsnames]{xcolor}
\usepackage{scrextend}
\usepackage{footnote}
\usepackage{color}
\usepackage{url}
\usepackage{bbding}
\usepackage{rotating}
\usepackage{tablefootnote}
\usepackage{listings}

\lstdefinelanguage{json}
{
    morestring=[b]",
    morestring=[d]'
}
   
\lstdefinelanguage{JavaScript}{
  keywords={break, case, catch, continue, debugger, default, delete, do, else, finally, for, function, if, in, instanceof, new, return, switch, this, throw, try, typeof, var, void, while, with},
  morecomment=[l]{//},
  morecomment=[s]{/*}{*/},
  morestring=[b]',
  morestring=[b]",
  sensitive=true
}

  \lstdefinelanguage{diff}{
    basicstyle=\ttfamily\small,
    morecomment=[f][\color{Green}]{+\ },
    morecomment=[f][\color{RedOrange}]{-\ },
  }
  \makeatletter
  \def\lst@makecaption{%
    \def\@captype{table}%
    \@makecaption
  }
  \makeatother

\newcommand{\devreplay}{\textsc{DevReplay}}

\newcommand{\rqone}{Can \devreplay~fix bugs better than state-of-the-art APR tools?}
\newcommand{\rqtwo}{What kind of bugs are \devreplay~effective?}
\newcommand{\rqthree}{Can \devreplay~suggest the same fixes as humans in code review?}
\newcommand{\rqfour}{Are generated patches accepted by project members?}

\begin{document}
\title{\devreplay: Automatic Repair \\with Editable Fix Pattern}



\author{Yuki Ueda         \and
        Takashi Ishio \and
        Akinori Ihara \and
        Kenichi Matsumoto
}
\maketitle



\begin{abstract}
Static analysis tools, or linters, detect violation of source code conventions to maintain project readability.
Those tools automatically fix specific violations while developers edit the source code.
However, existing tools are designed for the general conventions of programming languages.  These tools do not check the project/API-specific conventions.
We propose a novel static analysis tool \devreplay~that generate code change patterns by mining the code change history, and we recommend changes using the matched patterns.
Using~\devreplay, developers can automatically detect and fix project/API-specific problems in the code editor and code review.
Also, we evaluate the accuracy of \devreplay~using automatic program repair tool benchmarks and real software.
We found that \devreplay~resolves more bugs than state-of-the-art APR tools.
Finally, we submitted patches to the most popular open-source projects that are implemented by different languages, and project reviewers accepted 80\% (8 of 10) patches. 
\devreplay~is available on~\url{https://devreplay.github.io}.
\end{abstract}
\section{Introduction}

To reduce the time cost of maintenance, many software projects use static analysis tools.
Static analysis tools such as \textit{Pylint} automatically suggests that the source code changes to following the Python coding conventions~\cite{thenaultpylint}.
Also, many Automated Program Repair (\textbf{APR}) tools~\cite{Tool_elixir,Tool_avatar,Tool_par,APR_Survey}
have been presented that will fix source code not only for coding conventions, but also for the API misuse and run-time crashes.
However, static analysis tool users do not fix the around 90\% of tool warnings during the code editing~\cite{Panichella_SANER2015}.
One of the largest problems in the code editing is that the tool conversations are not stabled for each project.
One of the largest causes for this problem is that tool conventions are language general conventions, and not project/API-specific coding conventions.

Also, APR tools are believed to reduce the time costs of maintenance.
Template-based APR tools~\cite{Tool_phoenix, Tool_avatar,Tool_TBar} are recent trends in APR tools, but template-based APR tools require a lot of work for users to install and execute.
Tool users need to prepare complete test suites and collect change histories to make patches.
Additionally, users execute APR tools on the command line interface as an extra development process and must validate the APR tools output manually.

We present a static analysis tool, \textbf{\devreplay}~, that suggests the changes like APR tools.
However, unlike existing static analysis tools to optimize coding conventions for the project and API-specific styles, \devreplay~suggests source code changes using automatically generated code change patterns from a Git commit history.
Unlike existing APR tools, \devreplay~change suggestions are independent from test cases, and the required change history is less than 1 week.
Also, \devreplay~only uses a string match and it does not require type checks, syntax correctness, or context.
\devreplay~suggests source code changes whether or not the patch makes errors or not, and it does not depend on used programming languages.

We designed the \devreplay~suggestion algorithm based on template-based APR tools~\cite{Tool_elixir,Tool_avatar,Tool_par,APR_Survey}.
Table~\ref{tab:tools} shows tools comparison with~\devreplay, static analysis tools, and APR tools. \devreplay~has both feature from static analysis tools and APR tools.
To reduce tool user efforts needed for data collection, we implement a change distilling algorithm based on the research knowledge of APR tools~\cite{Nguyen_ICSE2019, Tool_avatar_pre}.
(1) First, most of the well-worked source code change templates are one hunk changes or one line changes~\cite{Nguyen_ICSE2019, Thung_MSR2012}.
To reduce the distilling area, \devreplay~generates source code change patterns from the one hunk changes, even though some bug-fixes need to change multiple locations. 
(2) Secondly, a half of change patterns appear within a month~\cite{Tool_avatar_pre}. 
\devreplay~learns such patterns from only recent (e.g. 1 day) git commits, while state-of-the-art APR tools used large data and filtered useful patterns.
\begin{table*}[t]
    \centering
    \caption{Comparison with~\devreplay, static analysis tools and APR tools}
    \begin{tabular}{l|l|l|l}
        & \devreplay &  Static analysis tools & Template-based APRs \\
        & (This paper) & (Subsection~\ref{subsec:sat}) & (Subsection~\ref{subsec:apr}) \\\hline
        Purpose & Prevent recent fix & Consist convention & Fix the bugs \\
        Fix contents & Recent similar problems & Coding style violation & Frequent templated fix \\
        Requirements & 10+ git commits & Nothing & 100K+ change history \\
        Customizing & Editing conventions &Editing conventions & Fixing test suites \\
        Executing & During Code Editing & During Code Editing & Maintaining process \\
        Language & 11 languages & One target language & C or Java  \\
        \hline
    \end{tabular}
    \label{tab:tools}
\end{table*}

In this paper, we make the following contributions.
\begin{itemize}
    \item We proposed a template-based static analysis tool, \devreplay, which focuses on the projects/API specific fixes that are not covered by current static analysis tools (Section~\ref{sec:motivatingeg}).  
    \item We present a method of generating a code change template format based on template-based APR tools. The template can be generated by recent 1 day git commits and it can be editable without AST knowledge (Section~\ref{sec:approach}).
\end{itemize}
\begin{figure*}[t]
    \centering
    \includegraphics[width=\linewidth]{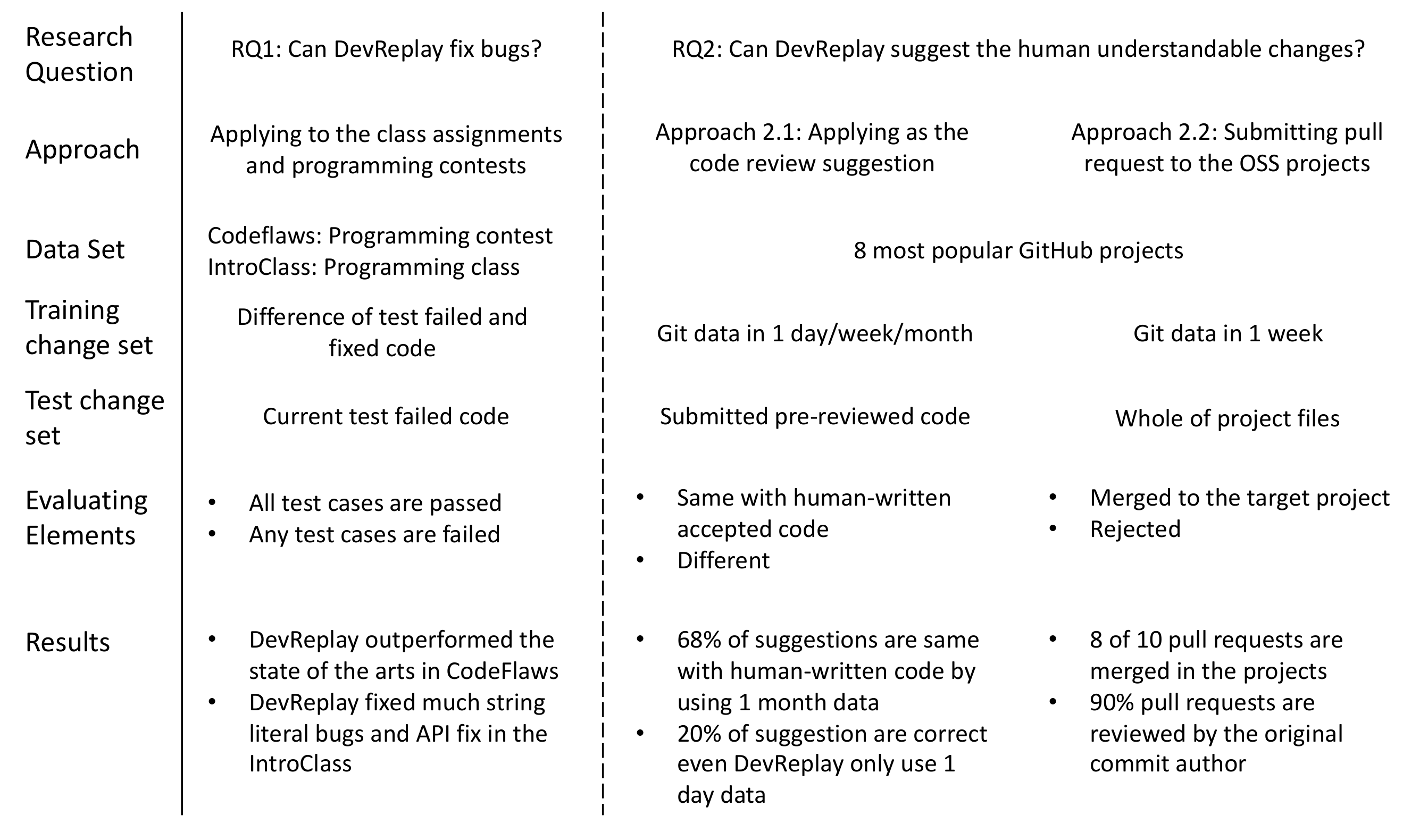}
    \caption{Research questions design and answers}
    \label{fig:overview}
\end{figure*}

In section~\ref{sec:evaluation},  We evaluate \devreplay~usefulness by answering the following research questions.
\begin{itemize}
    \item Benchmark evaluation (Subsection \ref{subsec:rqonetwo})
    \begin{itemize}
    \item RQ1: \rqone
    \item RQ2: \rqtwo
    \end{itemize}
    \item On the wild evaluation (Subsection \ref{subsec:onthewild})
    \begin{itemize}
    \item RQ3: \rqthree
    \item RQ4: \rqfour
    \end{itemize}
\end{itemize}
Figure~\ref{fig:overview} provides an overview of the our evaluation that has benchmark evaluations and in the wild evaluations.
In the benchmark evaluations, we simulate \devreplay~coverage by using two major APR benchmarks for C language.
In the RQ1 result, \devreplay~ outperformed to the state-of-the-art tools in the programming contest bugs.
Also in RQ2, \devreplay~solved especially for the string literal and API usage bugs that are not covered in the state-of-the-art APR tools.
In the wild evaluation, we evaluate the~\devreplay~usefulness for the real open source projects.
In the RQ3, we compare patches with the human-written code review changes.
As a result, we have suggested 68.6\% changes that are the same as the human-written changes by the 1-month patterns.
In the RQ4, we submit the patches to the original projects through the pull requests.
Finally, 8 of 10 submitted pull requests have already been accepted and merged by the development teams.

\section{Motivating Example}\label{sec:motivatingeg}
In this section, we show two motivating examples.
\devreplay~suggests the changes that solve the project as well as API and version-specific problems that are not covered by current static analysis tools.

The first example is a project-specific change in Listing~\ref{sr:motivating_lst}.
It includes two changes in the \textit{DefinitelyTyped} project, which is one of the most major open source projects.
The first change is a manual change by the developer of the module name, which fixes the project dependency by replacing the module name ``\texttt{NodeJS.Module}'' with ``\texttt {NodeModule}''.
The second change is automatically generated by \devreplay~reusing the manual change to maintain source code consistency.
This change is merged in the project by the original change author.

\begin{figure}[t]
    \begin{lstlisting}[caption=Motivating examples of the source code changes on the DefinitelyTyped project, label=sr:motivating_lst,language=diff]
// An example of the original DefinitelyTyped project change
- declare var require: NodeJS.Require;
- declare var module: NodeJS.Module;
+ declare var require: NodeRequire;
+ declare var module: NodeModule;

// An example of the represented change by DevReplay
// https://github.com/DefinitelyTyped/DefinitelyTyped/pull/41434/files
- interface Module extends NodeJS.Module {}
+ interface Module extends NodeModule {}
\end{lstlisting}
    \begin{lstlisting}[caption=Motivating examples of Language and API migrations, label=sr:motivating_python,language=diff]
// An example of Python2 to Python3 migration
- L = list(some_iterable)
- L.sort()
+ L = sorted(some_iterable)

// An example of the Chainer to PyTorch migration
- F.crelu($1, axis=1)
+ torch.cat((F.relu($1), F.relu(-$1)))
\end{lstlisting}
\end{figure}
The second motivating example is in Listing~\ref{sr:motivating_python}.
It shows two cases of typical API and language migrations fixes.
The first case shows the migration of the Python language version, and the second one shows the migration of ``Chainer'' packages.
These changes are supported by a migration guide~\cite{Chainer2PyTouch} and tools~\cite{2to3}, but using these migration is still costly.
An API user has to rewrite their source code manually or wait for a migration tool update for every language feature update.
To reduce the migration cost, \devreplay~users can make patterns by collecting migrated project histories, and can share these patterns with other users without the tool implementation knowledge.

Without \devreplay, both cases require the project, language, or API=specific knowledge that is not covered by existing static analysis tools.
Also, pattern authors can share their knowledge with other developers by sharing patterns files.

\section{Proposed method}\label{sec:approach}
\begin{figure*}[t]
    \centering
    \includegraphics[width=\linewidth]{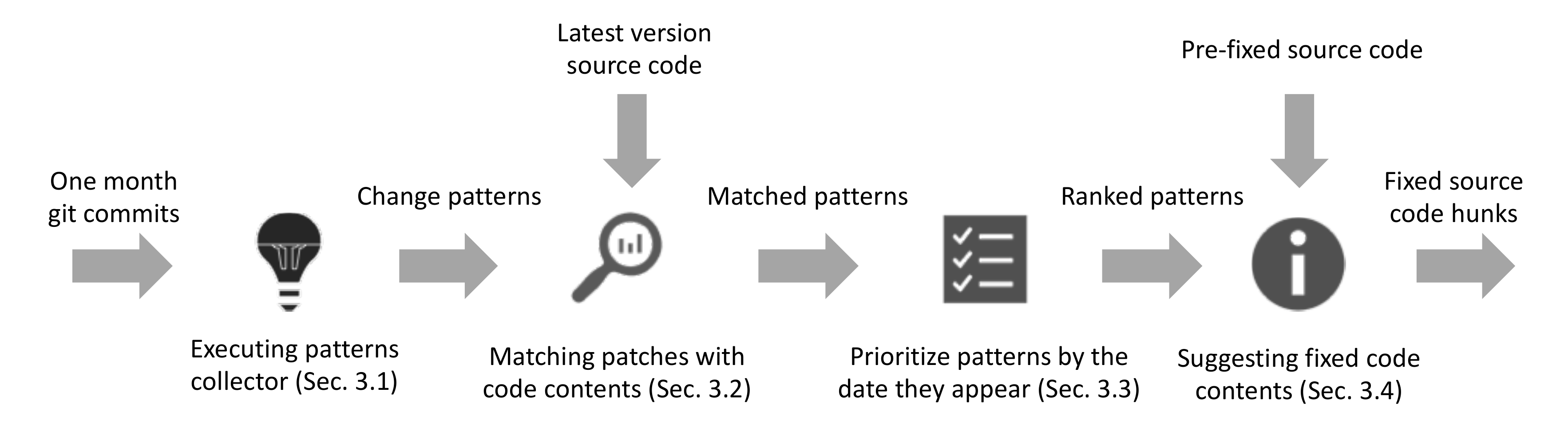}
    \caption{Overview of code fixing process with \devreplay}
    \label{fig:usecase_overview}
\end{figure*}
\devreplay~suggests code changes based on recent code changes to fix source code.
The tool works through four steps as illustrated in 
Figure~\ref{fig:usecase_overview}. 
\begin{enumerate}
    \item Extracting code change patterns from one month git commits (Subsection~\ref{sec:pattern_collecting}) 
    \item Matching patterns with code contents (Subsection~\ref{sec:matching})
    \item Prioritize patterns by change dates when they appear (Subsection~\ref{sec:prioritizing})
    \item Suggesting an applicable patch for each matched contents (Subsection~\ref{sec:suggesting}).
\end{enumerate}

\devreplay~generates human-readable code fix patterns that are written in TextMate snippet syntax~\cite{Textmate_snippets}.
TextMate snippets are widely used as to set the auto-complete function on several programming editors such as Vim and Visual Studio Code.

Users can manually edit the extracted patterns with familiar formats.
Using TextMate snippet syntax, developers can write change patterns not only in a numbered format, but also by just copying and pasting the real source code hunks.

\begin{figure}[t]
    \begin{lstlisting}[caption=Examples of the changes in TensorFlow project, label=sr:change_example,language=diff]
- _FORWARD_COMPATIBILITY_HORIZON = datetime.date(2020, 1, 4)
+ _FORWARD_COMPATIBILITY_HORIZON = datetime.date(2020, 1, 5)
\end{lstlisting}

    \begin{lstlisting}[caption=Examples of the change patterns from TensorFlow project, label=sr:pattern_example,language=json]
[
 {
  "repository": "tensorflow/tensorflow",
  "sha": "d0414a39f97fb99edc06a2943b4dba259d59fcf4",
  "author": "A. Unique TensorFlower",
  "created_at": "2020-01-05 18:02:30",
  "condition": [
   "$0 = $1.date($2, $3, 4)"
  ],
  "consequent": [
   "$0 = $1.date($2, $3, 5)"
  ],
  "abstracted": {
   "0": "_FORWARD_COMPATIBILITY_HORIZON",
   "1": "datetime",
   "2": "2020",
   "3": "1"
  }
 }
]
\end{lstlisting}
\end{figure}
Listings~\ref{sr:change_example} shows examples of changes from the TensorFlow project that changed the argument value from 4 to 5.
Listings~\ref{sr:pattern_example} shows patterns of Listings~\ref{sr:pattern_example} changes. In the pattern, common identifiers and numbers are abstracted such as \texttt{\$\{0:NAME\}} and \texttt{\$\{2:NUMBER\}}.
A developer can edit, add, or replace these patterns by confirming the generated change-patterns' snippets.

\subsection{Extracting code change patterns from git commit changes}\label{sec:pattern_collecting}
\devreplay~generates a TextMate snippet pair from the one month git commit history.
\devreplay generates patterns from commits within one month based on an existing study knowledge that shows half of the project changes are reused within one month~\cite{Nguyen_ICSE2019}.
We make the an abstract syntax tree (AST) from each revision by using ANTLR~\cite{Parr_ANTLR2013}, Ruby, and Go AST parser.
The style of hunks is depends on ANTLR syntax.
To imitate the human-written changes, we only changed the original grammar to detect only line breaks and white space changes.

\begin{figure}[t]
    \centering
    \includegraphics[width=\linewidth]{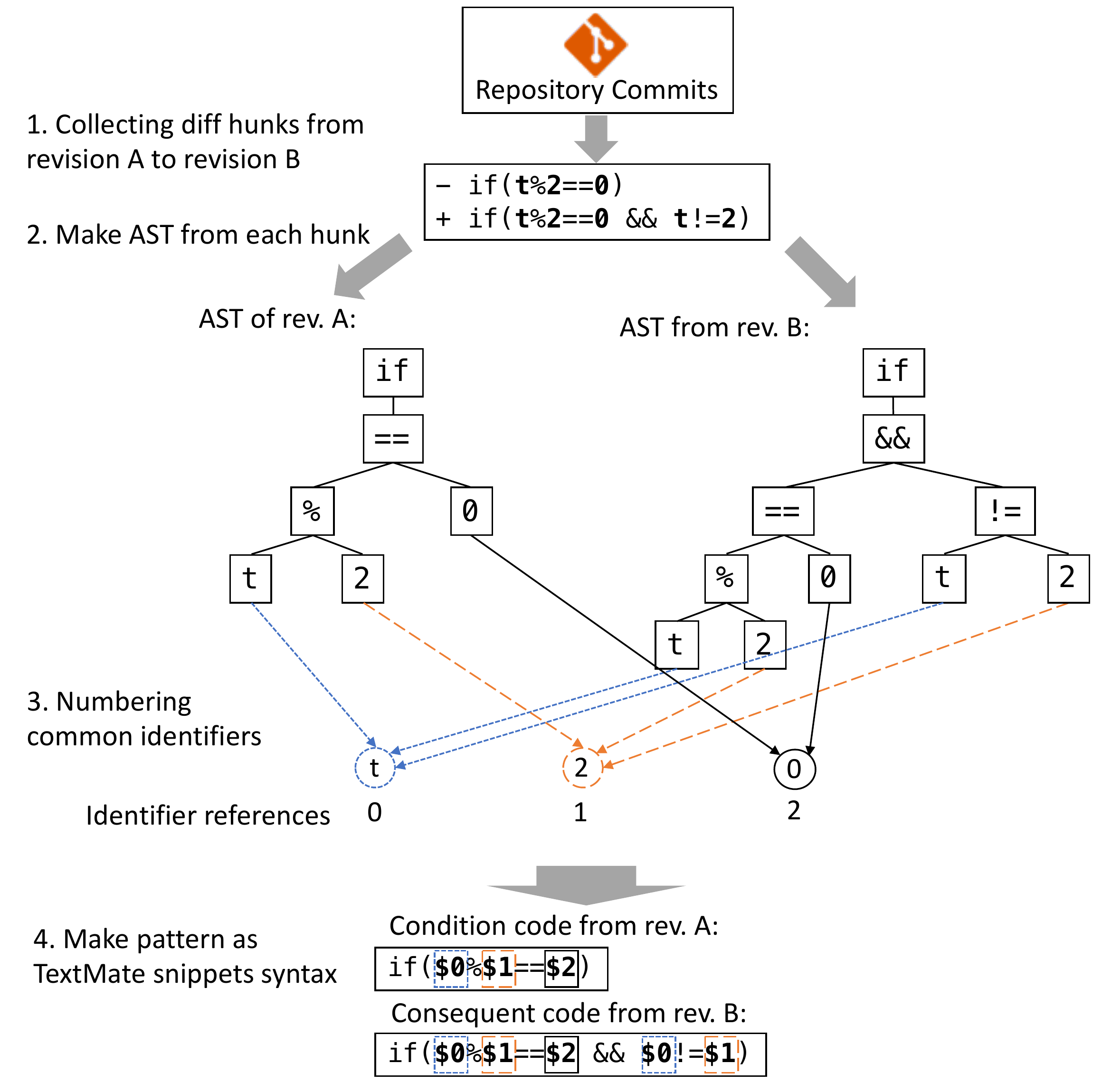}
    \caption{Change patterns-generating process by using the Git commits}
    \label{fig:learning}
\end{figure}

Fig~\ref{fig:learning} shows an overview of the \devreplay~pattern generating approach through the following process.
\begin{enumerate}
    \item Collecting differences from git repository revisions. In Figure~\ref{fig:learning}, \devreplay~collects the differences from revision A and revision B. In this process, we defined revision A as the pre-changed revision, and revision B as changed revision.
    \item Make two Abstract Syntax Trees (AST) from each revision hunks by using ANTLR.
    \item Number the tokens that are common to the two revision change hunks. The target tokens are identifiers, numbers, and string literals. This approach is based on an extracting coding idiom technique~\cite{Allamanis_TSE2018}. The original approach makes variable name references that is appeared in the one \texttt{for} loop idiom. Our approach has two differences, first, we refer to two different version hunks. Second, we detect not only the identifier, but also the string literal and number literal. In the figure~\ref{fig:learning},~\devreplay~numbers the identifier ``\texttt{t}'' as the ``\texttt{\$0}'', and numbers the literal ``\texttt{2}'' and ``\texttt{0}'' as the ``\texttt{\$1}'' and ``\texttt{\$2}'', respectively.
    \item Construct a change pattern as a TextMate snippet from revision A and revision B. 
    We define the snippet from revision A as the \textbf{condition pattern}, and the snippet from revision B as the \textbf{consequent pattern}. We replace common identifiers between both revisions to identifier IDs. For some differences that have multiple lines, \devreplay~generates the code change patterns collection. Finally, \devreplay~output collects patterns on the JSON file, ``\textit{devreplay.json}''.
\end{enumerate}
This change pattern of extracting scripts is available at\\~\url{https://github.com/devreplay/devreplay-pattern-generator}.

\subsection{Matching patches with code contents}\label{sec:matching}
In this process,~\devreplay~identifies source code hunks that are matched at any of the generated change patterns.
\devreplay~uses the numbered identifiers as the regular expressions. Unlike existing automatic program repairs~\cite{Tool_cvc4,Tool_avatar},~\devreplay~will not consider the token type during the patching process to extend the suggestable source code hunks.

\devreplay~detect matched source code by following steps.
\begin{enumerate}
    \item Convert condition code snippets to the JavaScript regular expression to detected numbered tokens. In this converting, we will not consider the tokens type. In the Listing~\ref{sr:pattern_example} ``condition'' element\\ ``\texttt{\$0 = \$1.date(\$2, \$3, 4)}'' becomes \\``\texttt{(?<token1>[\textbackslash w\textbackslash.]+) = (?<token2>[\textbackslash w\textbackslash.]+)\textbackslash .date((?<token3>[\textbackslash w\textbackslash.]+),\\ (?<token4>[\textbackslash w\textbackslash.]+), 4)}''
    \item Search source code hunks that are matched with condition snippets regular expressions from the target source code.
    \item Identify pattern that is matched with target source code. If more than two patterns are matched, we keep the both patterns.
\end{enumerate}

\subsection{Ordering patterns by when the date appears}\label{sec:prioritizing}

\devreplay~prioritizes the patterns recommendation.
According to existing studies, half of repeated fix are used in one month, and it is known that their distribution is more likely to occur as time approaches.
We add the the date of original change to the pattern, and the suggests the patterns in the order of the most recent dates.

\subsection{Suggesting fixed code contents}\label{sec:suggesting}

\devreplay~provides three user interfaces that include the command-line interface, GitHub code review bot, and code editor plugin.
In this section and evaluation, we assume the command-line interface to be the same as in the existing tools.
We introduce remained interface on the attachment~\ref{sec:attachment}.

Users modify the source code file on the command line interface that refers to the pattern file, provide the warnings or modify the fire for each matched code content.
To modify the source code, \devreplay~replaces the target source code identifier with the generated regular expression identifier using the replace function in JavaScript.

\section{Tool Evaluation}\label{sec:evaluation}

We evaluate \devreplay~usefulness by a benchmark evaluation and in the wild evaluation.
In this section, we answer the four research questions:
\begin{itemize}
    \item RQ1 (Subsection \ref{subsec:rqonetwo}): \rqone
    \item RQ2 (Subsection \ref{subsec:rqonetwo}): \rqtwo
    \item RQ3 (Subsection \ref{subsec:rqthree}): \rqthree
    \item RQ4 (Subsection \ref{subsec:rqfour}): \rqfour
\end{itemize}

\subsection{RQ1, 2: Benchmark evaluation with state of the art APR tools}\label{subsec:rqonetwo}

We evaluate the bug fix effectiveness of \devreplay~using the two major C benchmark data sets.

\subsubsection{Benchmark evaluation design}\label{sec:evaluation_design}

\textbf{Data set:}
We use the two benchmark sets, Codeflaws~\cite{Data_codeflaws} and IntroClass~\cite{Data_introclass} as the C language bug benchmark.
We compare the~\devreplay~benchmark result with the state-of-the-art 4 APR tools~\cite{Tool_angelix,Tool_cvc4,Tool_enum,Tool_semfix} that are evaluated in the existing study~\cite{Xuan2012}.
Both benchmarks consist of one source code file program.

The Codeflaws benchmark contains 3,902 defects collected from the Codeforces programming contest, with 6 categorized by bug types.
We select 665 bugs that are used as an evaluation study~\cite{Xuan2012}.
The selected bugs are from the ``replace relational operator'' (ORRN), the ``replace logical operator'' (OLLN), and the ``tighten or loosen condition'' (OILN) categories.
Listing~\ref{sr:codeflaw_defects} shows each bugs category source codes and fixed contents.
\begin{figure}[t]
    \begin{lstlisting}[caption=Examples of defect types from the Codeflaws data set used in our experiments, label=sr:codeflaw_defects,language=diff]
// An example of Replace relational operator (ORRN)
- if (sum > n)
+ if (sum >= n)

// An example of Replace logical operator (OLLN)
- if ((s[i] == '4') && (s[i] == '7'))
+ if ((s[i] == '4') || (s[i] == '7'))

// An example of 
// Tighten condition or loosen condition (OILN)
- if (t%2 == 0)
+ if (t%2 == 0 && t != 2)
\end{lstlisting}
\end{figure}
%

IntroClass consists of several hundred buggy versions of six different programs, written by students as homework assignments in a freshmen programming class.
The following shows the goals of each assignment:
\begin{itemize}
    \item smallest: Calculating the minimum value from the given 4 values.
    \item median: Calculating the median value from the given 3 values.
    \item digits: Calculating the given values' digits number.
    \item checksum: Counting the specific character from the strings
    \item grade: Calculating the student grade from the given values.
\end{itemize}
Each assignment is associated with two independent high-coverage test suites that are black-box test suites written by the course instructor, and white-box test suites generated by the automated test generation tool KLEE~\cite{Tool_Klee} on a reference solution.
Existing benchmark evaluation ignores the ``grade'' program. This program is related to the string type process that does not covered some target APR tools. However, we also solve the ``grade'' program to evaluate tool generalization performance.

\textbf{Approach:}
For the Codeflaws that have a time series ordered contest ID, we study one hypothesis that is the previous contest fixings can be reused for the current contest before submission.
For the IntroClass that does not has the time series data, we study the two hypotheses.
(1) The patterns from other class assignments fixes can be applied to the current assignment.
(2) The patterns from other students' fixes can be applied to the same assignment.

Using two data set, we evaluate the patch correctness by using four levels that are \textbf{Same}, \textbf{Success}, \textbf{Test failed}, and \textbf{No suggestion}.
\textbf{Same} is the any generated patches are the same with the human-written fixed patch. That cares about the white space but does not care about the code comment and new lines equality.
\textbf{Success} is all of the generated patches are different from the human-written patch, but any of them passed the all provided test cases.
\textbf{Test failed} is the any generated patch succeed the compile but all of them failed the more than one test case.
\textbf{No suggestion} is the all of the generated patches failed the compile, it is also include the case of~\devreplay~could not generate any patch.
We define \textit{Same} and \textit{Success} patches are \textit{fixed patches}, and \textit{Test Failed} and \textit{No suggestion} patches are \textit{failed patches}.
\begin{figure}
    \centering
    \includegraphics[width=\linewidth]{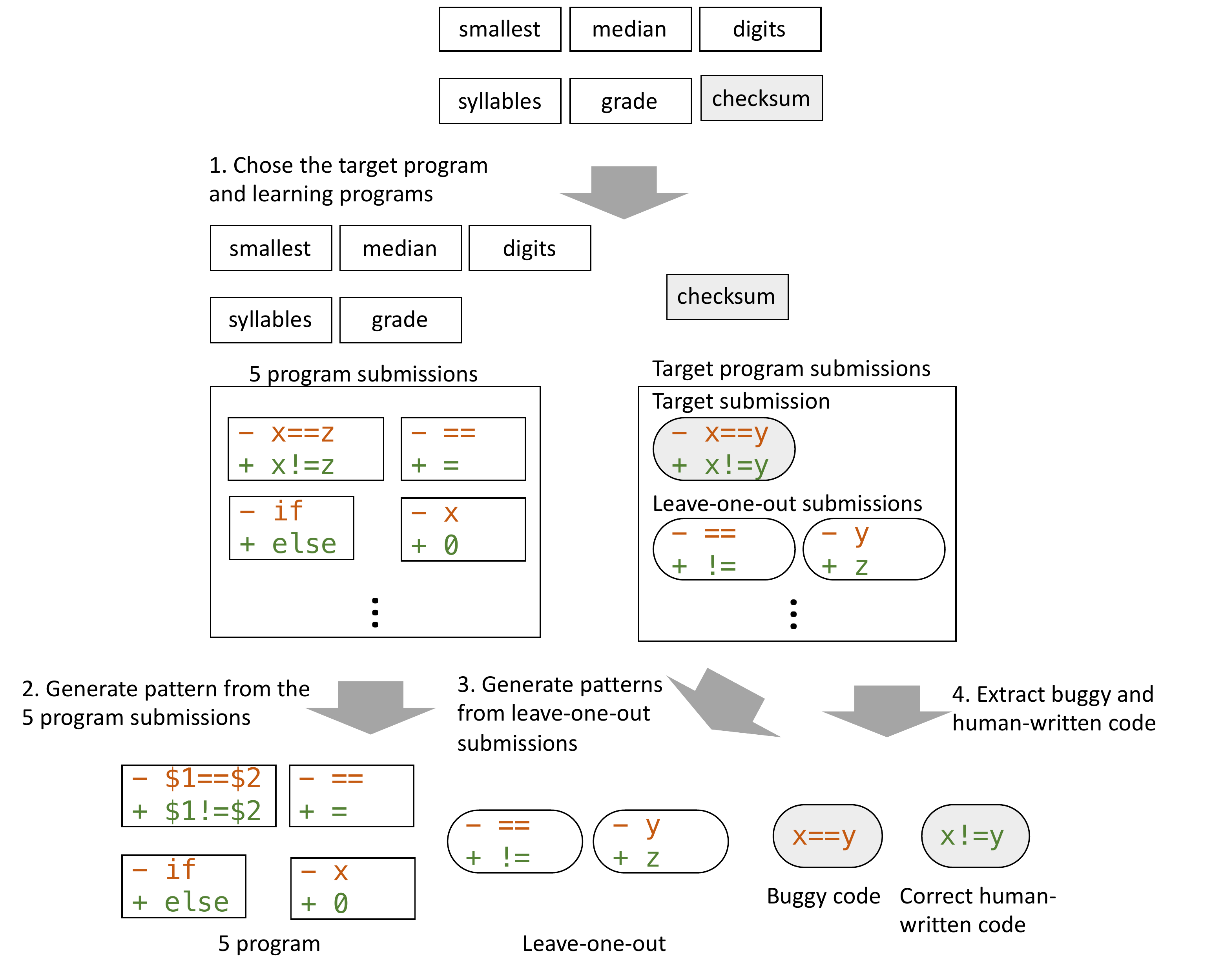}
    \caption{Process of parsing the IntroClass data set}
    \label{fig:introclass_method}
\end{figure}

To generate change patterns for CodeFlaws in time series, we use the contest IDs, which are ordered in the contest dates.
\devreplay~generates change patterns from the past programming contests that have small contest ID rather than a target programming contest.
In this evaluation,~\devreplay~suggests the patches order by the latest contest patterns.
Figure~\ref{fig:codeflaws_method} provides an overview for following evaluation steps:
\begin{figure}
    \centering
    \includegraphics[width=\linewidth]{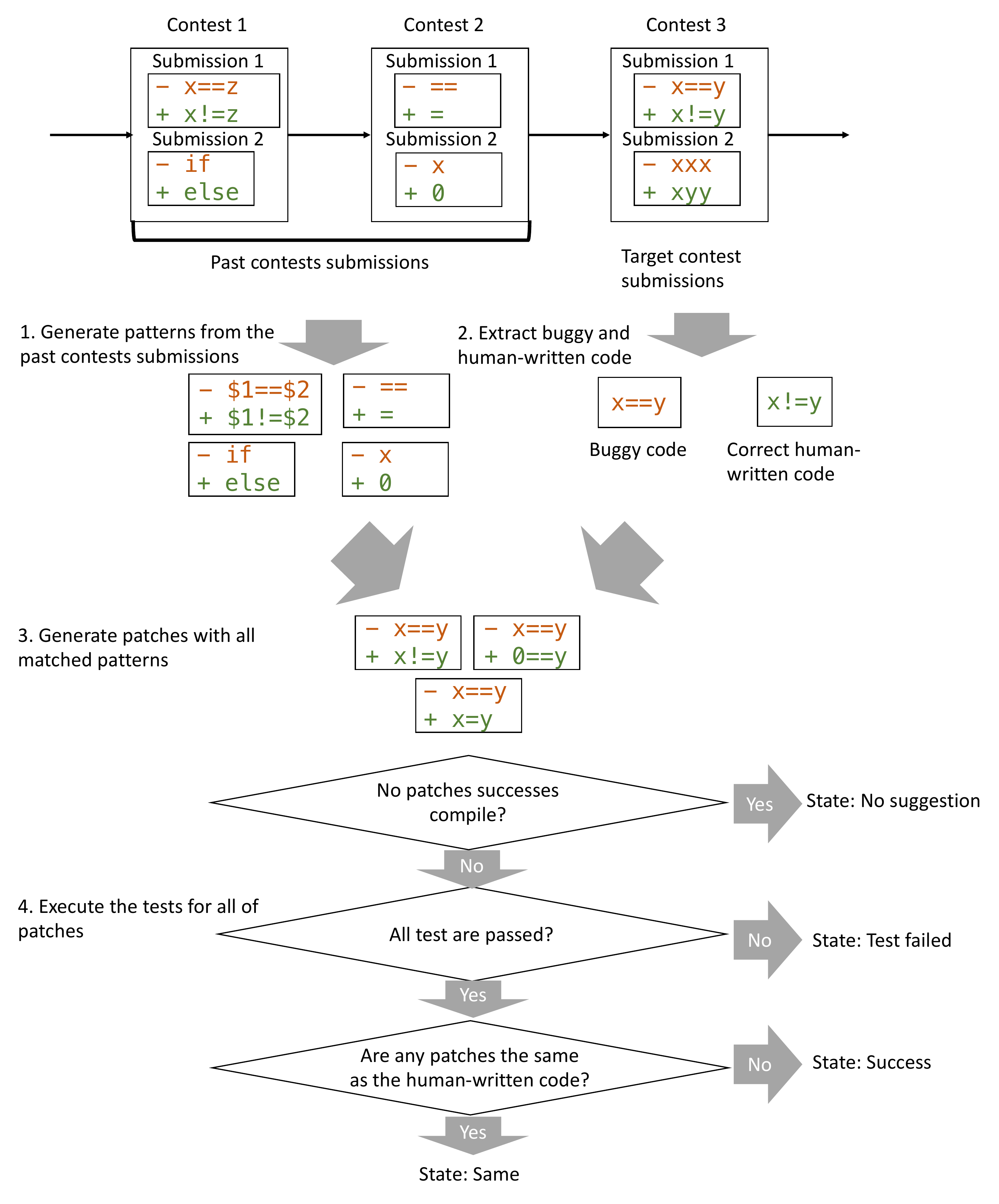}
    \caption{Overview for evaluation on Codeflaws data set}
    \label{fig:codeflaws_method}
\end{figure}
\begin{enumerate}
    \item Generating code change patterns from past programming contest submissions. In Figure~\ref{fig:codeflaws_method}, we extract 4 patterns that include a numbered pattern. 
    \item Extracting buggy and human-written code from target contest submissions.
    \item Generate patches based on matched patterns. In this case, if~\devreplay cannot generate any patches or if all patches fail to compile, the state will be represented by \textit{No suggestion}.
    \item Execute tests for all generated patches if the state is not a \textit{No suggestion}. From the test result and differences with the human-written code, we identify the result as \textit{Test failed}, \textit{Success}, or \textit{Same}. 
\end{enumerate}
In this evaluation, \devreplay~does not suggest any patterns to the small ID contests that have few past contests.
In a real environment, this limitation can be solved by transferring the patterns from other projects.

For the IntroClass, to answer the two hypotheses, we use the three validating method.
First, we generate the patterns from the submissions of the same assignment by using the leave-one-out (LOO) cross validation~\cite{allamanis2014learning}. We call the patterns the \textbf{LOO patterns}. Using the LOO patterns, we evaluate the tools' specialization performance.
Second, we generate the patterns from 5 program assignment submissions that are not targeted assignment. We call the patterns the \textbf{5 program patterns}. Using 5 program patterns, we evaluate the tools' generalization performance.
Finally, we use both of the LOO and the 5 program patterns to evaluate two method' coverage differences.
Even through source code is labeled as the buggy, some source code files are worked without any changes.
In the IntroClass especially, some patterns do not change the source code file. We remove these patterns from our investigation.

Figure~\ref{fig:introclass_method} shows the overview of the evaluation method on IntroClass.
In this evaluation,~\devreplay~generate change patterns for each program from the other 5 programs and leave one out.

We execute the following evaluation process for IntroClass
\begin{enumerate}
    \item Choose 5 learning set programs and 1 target program. In Figure~\ref{fig:introclass_method}, we chose the ``checksum'' program as the target program, and chose ``digits'',  ``median'', ``smallest'' and ``syllables'' programs as learning set programs.
    \item Generate patterns from buggy and human-written fixed code in the 5 programs submissions.
    \item Generate patterns from the non-targeted (LOO) submissions in target programs.
    \item Extract buggy and human-written code from the target program submission.
\end{enumerate}
Next, since the remaining evaluation is the same as Codeflaws, we omit these process from Figure~\ref{fig:introclass_method}
\begin{enumerate}
    \item[5.] Generate patches and compare with human-written code. In this case, If~\devreplay can not find the patches or all patches fail compile, the state represents \textit{No suggestion}
    \item[6.] Execute tests for all of the patches if the state does not match \textit{No suggestion}. From the test results and the difference with the human-written code, we can identify the result as \textit{Test failed}, \textit{Success}, or the \textit{Same}. 
\end{enumerate}

\subsubsection{Benchmark evaluation result}\label{sec:evaluation_result}

\begin{table*}[t]
    \centering
    \caption{651 Codeflaw bug fix for each APR tools}
    \begin{tabular}{l|rr}
Tool &Same/Success & Test failed/No suggestion \\\hline
Angelix &  81 (12.4\%) & 570 \\
CVC4 & 91 (14.0\%) & 560 \\
Enum & 92 (14.1\%) & 559 \\
Semfix & 56 ( 8.6\%) & 595 \\\hline
DevReplay (1 patterns) &  55 ( 8.0\%) & 596 \\
DevReplay (3 patterns) &  90 (13.4\%) & 561 \\
DevReplay (5 patterns) &  \textbf{101 (14.9\%)} & 550 \\
DevReplay (All patterns) &  \textbf{136 (20.9\%)} & 515 \\
    \end{tabular}
    \label{tab:codeflaws}
\end{table*}

We discuss the result and answering the two findings for each benchmark data set.
\begin{enumerate}
    \item RQ1: \rqone
    \item RQ2: \rqtwo
\end{enumerate}

First, we compare the~\devreplay~bug fix coverage in Codeflaws. 
Table~\ref{tab:codeflaws} shows a comparison result with the state-of-the-art APR tools for Codeflaws.
Table~\ref{tab:codeflaws} shows the number of fixed bugs for the number of the suggested patterns.
\devreplay~outperformed state-of-the-art by only referring to the latest five patterns that match the conditions, without reference to all of the patterns.

\begin{table*}[t]
    \centering
    \caption{IntroClass bug fix accuracy for DevReplay and APR tools}
\begin{tabular}{lrrrrrrr}
Black-test &&\multicolumn{6}{c}{\# of same/success fix (\# of same fix): \% of succeed coverage}     \\\hline
Program   & \#  & DevReplay       & \#  & Angelix    & CVC4       & Enum       & Semfix     \\
smallest  & 68  & 8 (0)   & 56  & 37 & 39& 29& 45 \\
  &   &11.8\%   &   & 66.1\% & 69.6\% & 51.8\% &80.4\% \\
median    & 64  & 1 (0)    & 54  & 38 & 28 & 27 & 44 \\
    &   & 1.6\%    &   & 70.4\% & 51.9\% & 50.0\% & 81.5\% \\
digits    & 59  & 0 (0)    & 57  & 6  & 4   & 3   & 10 \\
    &   & 0.00\%    &   & 10.5\%  & 7.0\%   & 5.3\%   & 17.5\% \\
syllables & 39  & 18 (1)  & 39  & 0   & 0   & 0   & 0   \\
 &   & \textbf{46.1\%}  &   & 0.0\%   & 0.0\%   &0.0\%   & 0.0\%   \\
checksum  & 18  & 0 (0)    & 19  & 0  & 0   & 0   & 0   \\
  &  & 0.00\%    &   & 0.0\%   & 0.0\%   & 0.0\%   & 0.0\%   \\
grade     & 96  & 14 (9)  & --  & --         & --         & --         & --         \\
     &  &  \textbf{14.6\%}  &   & --         & -- \\
total    & 344 & 41 (10) & 225 & 81 & 71 & 59 & 99 \\
    &  & 11.9\% &  & 36.0\% & 31.6\% & 26.2\% & 44.0\% \\
\\
White-test &&\multicolumn{6}{c}{\# of same/success fix (\# of same fix): \% of succeed coverage}    \\\hline
program   & \#  & DevReplay  & \#  & Angelix    & CVC4       & Enum       & Semfix     \\
smallest  & 49  & 4 (0)    & 41  & 37 & 37 & 36 & 37 \\
  & 49  & 8.2\%    &   & 90.2\% &  90.2\% &  87.8\% & 90.2\% \\
median    & 52  & 1 (0)    & 45  & 35 & 36 & 23 & 38 \\
    & 52  & 1.9\%    &   & 77.8\% & 80.0\% & 51.1\% & 84.4\% \\
digits    & 94  & 0 (0)    & 90  & 5   & 2   & 2   & 8   \\
   & 94  & 0.0\%    &  & 5.6\%   & 2.2\%   & 2.2\%   &  8.9\%   \\
syllables & 46  & 23 (1)`  & 42  & 0   & 0   & 0   & 0   \\
 &   & \textbf{50.0\%}  &   &  0.0\%   & 0.0\%   &  0.0\%   & 0.0\%   \\
checksum  & 30  & 0 (0)    & 31  & 0   & 0  & 0   & 0   \\
  &   &  0.00\%    &  & 0.0\%   & 0.0\%   & 0.0\%   &  0.0\%   \\
grade     & 95  & 14 (9)  & --  & --         & --         & --         & --         \\
    & 95  & \textbf{14.7\%}  &  & --         & --         & --         & --         \\
total     & 366 & 42 (10) & 249 & 77 & 75 & 61 & 83 \\
     & & 11.5\% &  &  30.9\% & 30.1\% & 24.5\% & 33.3\%\\
\end{tabular}
\label{tab:introclass}
\end{table*}

Next, we discuss the bug fix coverage and fixed bug kinds in  the IntroClass.
Table~\ref{tab:introclass} shows the number of fixed bugs for each program assignment.
The difference in the number of bugs is due to the version of the data set and the filtering in the existing study.
Therefore, the authors compare \devreplay~with existing tools by coverage, and not by the absolute number.
As a result, we found that IntroClass provided greater coverage than other tools for \textit{syllables} and \textit{grades} not covered by other tools.
The format of the character string can be found in other assignments and in other submissions with similar modifications that are effective for \devreplay.
However, numerical calculations such as \textit{digit} are difficult to correct with \devreplay, which makes corrections without validating behavior.
\\
\fbox{
\begin{tabular}{l}RQ1 Answer: \\
\devreplay~outperformed in the Codeflaw\\
(programming contest) \\
data set that only focused on similar bugs.\\
Also, \devreplay~worked in IntroClass \\
(class assignments) bugs \\
that are not covered by state-of-the-art APR tools.
\end{tabular}
}

\begin{table*}[t]
    \centering
    \caption{651 Codeflaw bug fix result for each bug kind}
    \begin{tabular}{l|rrrrr}
Bug Kind &  Total & Same &  Success &  Test failed &  No suggestion \\\hline
OILN     &    325 &    0 ( ~0.0\%) &       52 (16.0\%) &          152 &             121 \\
OLLN     &     18 &    0 ( ~0.0\%) &        1 (~5.6\%) &            6 &              11 \\
ORRN     &    308 & 46 (14.9\%) &       37 (12.0\%) &          127 &              98 \\
\hline
Total & 651 & 46 (7.1\%) & 90 (13.8\%) & 285 (43.8\%) & 230(33.8\%)\\
&     &  \multicolumn{2}{c}{136 (20.9\%)} & \multicolumn{2}{c}{515 (79.1\%)} \\
    \end{tabular}
    \label{tab:codeflaw_kind}
\end{table*}
To answering RQ2, Table~\ref{tab:codeflaw_kind} shows the fixed bug kinds that are shown in Listing~\ref{sr:codeflaw_defects}.
Problems solved by~\devreplay~have a bias for the bug kinds or programs.
For the Codeflaws, \devreplay~solved the ORRN (Replace relational operator) by the same fix as human-written.
Also, OILN (Tighten condition or loosen condition) passedthe  test cases.
However, most OLLN (Replace logical operator) are not fixed.
In one of the causes, the OLLN learning set is than other in the two bug kinds.
In the ORRN and OILN, most failed states are \textit{Test failed}, but OILN failed as \textit{No suggestion}.

\begin{table}[t]
    \centering
    \caption{IntroClass bug fix accuracy for~\devreplay~training data set}
\begin{tabular}{lrrrr}
Black-test &     &               &               &\\\hline
Program   & \#  & Leave-one-out  & 5 programs     & LOO+5 programs  \\
smallest  & 68  & 6(0): ~8.8\%   & 2(0): ~2.9\%   & 8(0): 11.8\%   \\
median    & 64  & 1(0): ~1.6\%   & 1(0): ~1.6\%   & 1(0): ~1.6\%    \\
digits    & 59  & 0(0): ~0.0\%   & 0(0): ~0.0\%   & 0(0): ~0.0\%    \\
syllables & 39  & 4(0): 10.3\%  & 17(0):\textbf{43.6\%} & 18(1): 46.2\%  \\
checksum  & 18  & 0(0): ~0.0\%   & 0(0): ~0.0\%   & 0(0): ~0.0\%    \\
grade     & 96  & 14(9):\textbf{14.6\%} & 0(0): ~0.0\%   & 14(9): 14.6\%  \\
total     & 344 & 25(9): ~7.3\%  & 20(0): ~5.8\%  & 41(10): 11.9\% \\
\\
White-test &     &               &               &\\\hline
program   & \#  & Leave-one-out  & 5 programs     & LOO+5 programs  \\
smallest  & 49  & 4(0): ~8.1\%   & 1(0): ~2.0\%   & 4(0): ~8.1\%    \\
median    & 52  & 1(0): ~1.9\%   & 0(0): ~0.0\%   & 1(0): ~1.9\%    \\
digits    & 94  & 0(0): ~0.0\%   & 0(0): ~0.0\%   & 0(0): ~0.0\%    \\
syllables & 46  & 4(0): ~8.7\%   & 23(0): \textbf{50.0\%} & 23(1): 50.0\%  \\
checksum  & 30  & 0(0): ~0.0\%   & 0(0): ~0.0\%   & 0(0): ~0.0\%    \\
grade     & 95  & 14(9):\textbf{14.7\%} & 0(0): ~0.0\%   & 14(9): 14.7\%  \\
total     & 366 & 23(9): ~6.3\%  & 24(0): ~6.6\%  & 42(10): 11.5\%\\
\label{tab:introclass_approach}
\end{tabular}
\end{table}

Table~\ref{tab:introclass_approach} shows the coverage for Leave-one-out(LOO) pattern coverage, 5 program pattern coverage and sum of their sum.
LOO worked for \textit{grade} and 5 programs worked for \textit{syllables}.
In addition, the sum of both shows that the ranges modified by LOO and the 5 programs for \textit{smallest}, \textit{grade}, and \textit{syllables} do not overlap.
We designed LOO and 5 programs to evaluate the performance of specialization and generalization.
When applying \devreplay~to assignments, it is better to use both cases as much as possible.

\begin{table}[t]
\centering
    \caption{Patterns that appear the most in the IntroClass}
    \begin{tabular}{lrrr}
Pattern in Leave one out                               & White& Black  &Both\\\hline
FixPrintString       & 13    & 13    & 26    \\
FixBoundary          & 6     & 11    & 17    \\
AddSpace             & 3     & 1     & 4     \\
Others               & 1     & 0     & 1     \\
\\
Pattern  in 5 programs                            & White& Black  &Both\\\hline
AlternativeFunctionCall                 & 13    & 13    & 26    \\
AddReturnParentheses & 9     & 3     & 12    \\
AddSpace             & 2     & 1     & 3     \\
AddDefinition        & 0     & 2     & 2     \\
FixBoundary          & 0     & 1     & 1    
\end{tabular}
    \label{tab:introclass_patterns}
\end{table}
Next, to identifying the commonly fixed bug kinds, we manually labeled the patterns that are used in the \textit{Same} and \textit{Success} bug fixes.
Table~\ref{tab:introclass_patterns} show the most frequently used patterns in the IntroClass evaluation.
Also, Listing~\ref{sr:introclass_example} shows a concrete example for each pattern that appears.
Some changes do not have a behavioral impact (e.g. \texttt{AddSpace}). These patterns have bugs that appeared and these patterns are worked without any changes.
Existing studies found that the most frequently appearing change content is the \texttt{Null-pointer-check}~\cite{Monperrus_ICSE2014}.
Unlike these studies,~\devreplay~suggests changes that are related to string literal and API misuse.
In the LOO validation, \texttt{FixPrintString} is the most frequently appearing pattern.
22 of 26 \texttt{FixPrintString}s target entire string literal to adjusting the test suites.
However, the remaining \texttt{FixPrintString} only focuses on small changes such as ``\texttt{Stdent}'' to ``\texttt{Student}''.
Throughout the development with \devreplay, users can find and prevent the common bugs that are not covered by the existing tools.
In the 5 programs, \devreplay~suggests \texttt{OFFN} (API misuse) that appeared on the cross in different programs that fix the standard input size.

\fbox{
\begin{tabular}{l}RQ2 Answer: \\
\devreplay~fixed string literal and API misuse\\
bugs, that are not covered by state-of-the-art\\
APR tools.
\end{tabular}
}

\begin{figure}[t]
    \begin{lstlisting}[caption=Examples of the fixed C source code in IntroClass, label=sr:introclass_example,language=diff]
// FixPrintString
- printf(\"Student has an F grade\\n\");
+ printf(\"Student has failed the course\\n\");

// FixBoundary
- if ($0 > $1)
+ if ($0 >= $1)
//or
- char $0[21];
+ char $0[20];

// AddSpace
- int main(){
+ int main (){

// OFFN (AlternativeFunctionCall)
- scanf(\"%s\", $0);
+ fgets($0, sizeof($0), stdin);
//or
+ fgets($0,256, stdin);

// AddReturnParentheses
- return $0;
+ return ($0);

// AddDefinition
- int $0, $1;
+ int $0, $1, x;
//or
- int $0;
+ int $0 = 0;
\end{lstlisting}
\end{figure}

\subsection{RQ3, 4: On the wild evaluation}\label{subsec:onthewild}

The authors answer this research question through the two hypothesis and approaches in major open source projects.
Hypothesis 2.1: Code review costs can be reduced by automatically using the past code changes.
By using the~\devreplay~GitHub application implementation, project team members can extend the developers knowledge to the project knowledge.
Hypothesis 2.2: Developers can automatically fix what they have missed to modify their code files.
When the code change is not applied some project files,~\devreplay~can suggest to developers how to fix the unfixed code.

\subsubsection{RQ3: \rqthree}\label{subsec:rqthree}

\textbf{Data set:}
We use the open source project code review works as the target data set.
Also, we chose the 10 highest contributors target projects~\cite{Octverse} in Table~\ref{tab:review_projects}.
From these projects, our targeted 7 of 10 projects that are followed two conditions: (1) The projects are source code projects, which means that more than half of files from the projects are not document files. With this condition (2) projects pull request commits are traceable. The Facebook/react-native project uses the particular pull request process that closes the all pull requests even if the suggestions are accepted.
We generate patterns from the 2018s to 2019s commit history from these projects.
For the change pattern, we defined the submitted code as the condition code, and defined merged code as the consequent code.

\begin{table}[t]
\centering
\caption{Target OSS projects. RQ3 and RQ4 focus on the source code projects, and RQ3 focuses on applying the traceable code review project.}
\begin{tabular}{llrr}
&&   \# of&   \\
Project & Main Language &Contributors& Evaluate on \\\hline
VS Code & TypeScript & 19.1k & RQ3 and RQ4 \\
Azure-docs & Markdown &14k & Nothing \\
Flutter & Dart & 13.0K & RQ3 and RQ4 \\
First-contributions & Markdown &11.6k&Nothing\\
React Native & Python & 11.6K & RQ4\\
Kubernetes & Go & 11.6K & RQ3 and RQ4\\
TensorFlow & C++ & 9.9K &  RQ3 and RQ4\\
DefinitelyTyped & TypeScript & 6.9K & RQ3 and RQ4\\
Ansible & Python & 6.8K& RQ3 and RQ4\\
Home Assistant & Python & 6.3K& RQ3 and RQ4 \\
\end{tabular}
\label{tab:review_projects}
\end{table}

\begin{figure}
    \centering
    \includegraphics[width=\linewidth]{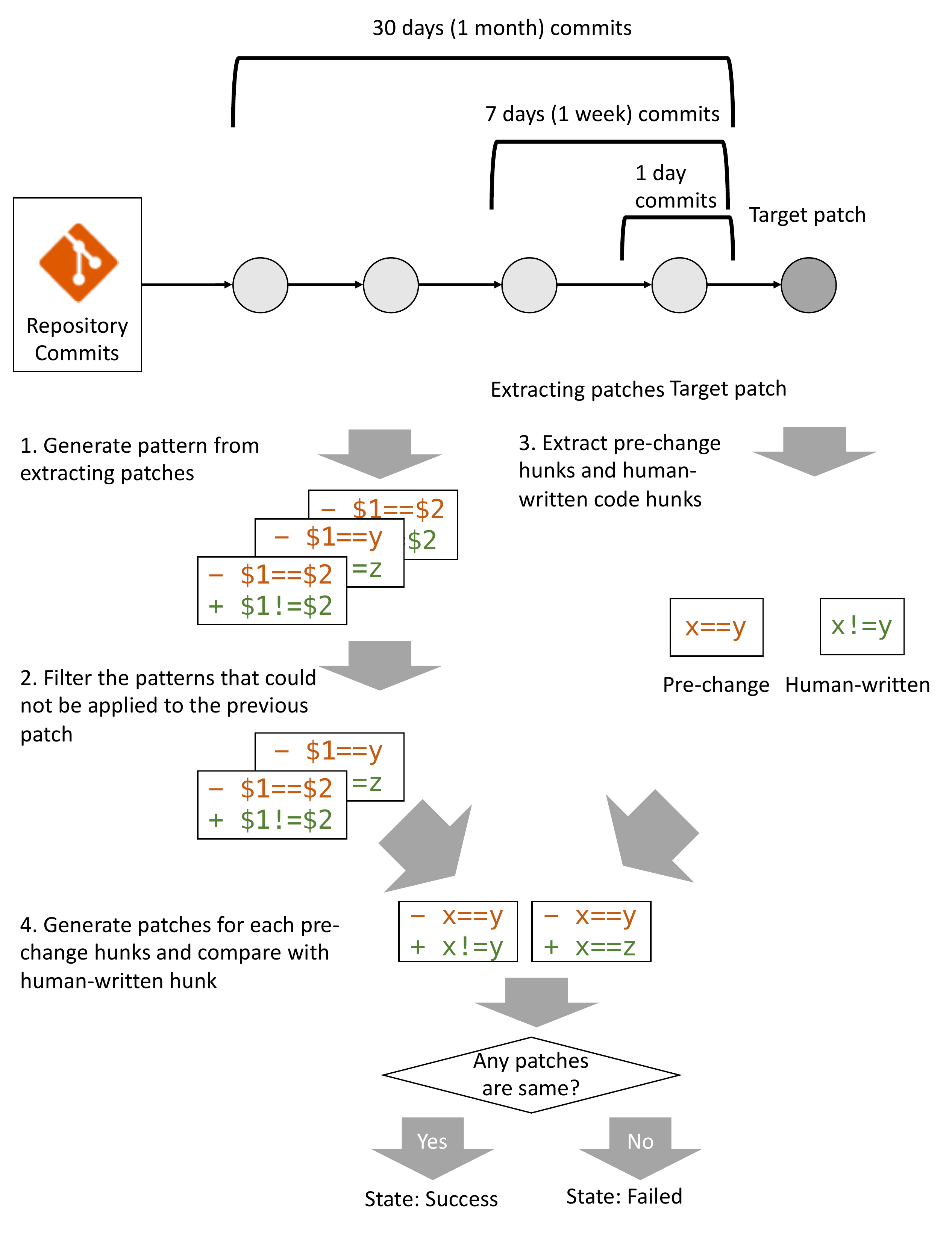}
    \caption{Overview for evaluation of git base project data set}
    \label{fig:git_method}
\end{figure}

\textbf{Approach: }
We made the patch and evaluated it using the following process.
We referred to the existing study that said half of the repeated code changes appeared with in a month, and we used the one day/week/month patterns to reduce the learning data set~\cite{Tool_avatar_pre}.
Figure~\ref{fig:git_method} provides an overview for extracting a git base project data set and evaluating the generated patches.

\begin{enumerate}
    \item Generate change pattern from target project git history or code review history. In this manner, we use the patterns within the time period they appeared. 
    \item If the patterns fail to applying to the past changes, we removed those patterns from the suggestion.
    \item Extract pre-change hunks and human-written code hunks from target code review data set.
    \item Generate patches for each pre-change hunk and compare with human-written hunk. If any patches are the same as the human-written hunk, the state is ``'Success''; if not, the state is ``Failed''.
\end{enumerate}

\textbf{Results:}
Table~\ref{tab:rq3result} shows the suggest precision for each project and time period.
For the best results, Flutter projects that the precision for a 30-days commit history is 68.63\%.
From the entire result, we found the large data set project (e.g. flutter and home-assistant) and large time period precision have high precision and fixed contents.
Our goal was to get the trending patterns. However, tool users can find the permanent patterns from the long term.

\fbox{
\begin{tabular}{l}RQ3 Answer: \\
\devreplay~worked by $<$ 40\% precision in code review\\
even if it only used the 1 day commit history.
\end{tabular}
}

\begin{table*}[t]
\centering
\caption{Accuracy of code review suggestions for each hunks}
\begin{tabular}{l|rrrr}
&& \multicolumn{3}{c}{Success/Suggested (Accuracy)}\\
Project    & \#     & 1 day patterns& 7 days patterns   & 30 days patterns                 \\\hline
VS Code&687,340&700 / 3,167&1,753 / 6,886 &2,379 / 8,544\\
&&22.1 \%&25.5 \%& 27.8 \%\\
Flutter   & 245,974 & 217 / 605  & 893 / 1,565 & 1,763 / 2,569 \\
&&35.9 \%&57.1 \% &\textbf{68.6 \%} \\
Kubernetes & 53,346  & 114 / 754  & 753 / 2,378 & 1,426 / 3,956  \\
&&15.1 \%& 31.7 \%&36.1 \%\\
TensorFlow&1,039,298&1,973 / 4,864 &6,133 / 11,273 &8,920 / 14,930 \\
&&\textbf{40.6 \%}&54.4 \%&59.7\%\\
DefinitelyTyped &1,219,731&662 / 2,389 &2,302 / 5,036 &5,037 / 8,072 \\
&&27.7 \%&45.7 \%& 62.4 \%\\
Ansible       & 20,503  & 9 / 586      & 154 / 1,080  & 295 / 1,530   \\
&&1.5 \%& 14.3 \%&19.3 \%\\
Home Assistant & 235,966 & 184 / 1,096  & 1,575 / 3,598  & 3,594 / 6,201\\
&&16.8 \%&43.8 \%& 58.0 \%

\end{tabular}
\label{tab:rq3result}
\end{table*}

\subsubsection{RQ4: \rqfour}\label{subsec:rqfour}

We conducted a live study to evaluate the effectiveness of fix patterns to maintain consistency in open source projects.
In this study, we used \devreplay~for the source code change suggestion.

\textbf{Data set:}
Table~\ref{tab:submitted_projects} shows the target projects.
As in RQ3, we used the 8 source code project commit history from the 10 most popular projects.
Unlike the predicting evaluation on RQ3, some code change suggestions may be too old to be reviewed by project members.
We only use the one-week git commits on the target project to detect the change patterns that are still fresh in project members' memory.

\begin{table*}[t]
\centering
\caption{Submitted pull requests and used time span}
\begin{tabular}{llrrl}
Project & Language & Pattern collecting span & State  \\\hline
VS Code & TypeScript & 2020/01/01--2020/01/07  & Merged~\footnotemark\\
&&& Closed~\footnotemark \\
Flutter & Dart & 2020/01/28--2020/02/04& Closed~\footnotemark\\
React Native & JavaScript/Java/C++ & 2020/01/17--2020/01/24& Merged~\footnotemark \\
Kubernetes & Go & 2020/01/29--2020/02/05& Merged~\footnotemark \\
TensorFlow & C++/Python & 2020/01/01--2020/01/07& Merged~\footnotemark\\
DefinitelyTyped & TypeScript & 2020/01/01--2020/01/07& Merged~\footnotemark \\
Ansible & Python & 2020/01/01--2020/01/07 & Merged~\footnotemark \\
&&& Merged~\footnotemark \\
Home Assistant & Python & 2020/02/06--2020/02/13& Merged~\footnotemark \\
\end{tabular}
\label{tab:submitted_projects}
\end{table*}
\footnotetext[1]{vscode\#87709: \url{https://github.com/microsoft/vscode/pull/87709}}
\footnotetext[2]{vscode\#88117: \url{https://github.com/microsoft/vscode/pull/88117}}
\footnotetext[3]{flutter\#50089: \url{https://github.com/flutter/flutter/pull/50089}}
\footnotetext[4]{react-native\#27850: \url{https://github.com/facebook/react-native/pull/27850}}
\footnotetext[5]{kubernetes\#87838: \url{https://github.com/kubernetes/kubernetes/pull/87838}}
\footnotetext[6]{tensorflow\#35600: \url{https://github.com/tensorflow/tensorflow/pull/35600}}
\footnotetext[7]{DefinitelyTyped\#41434: \url{https://github.com/DefinitelyTyped/DefinitelyTyped/pull/41434}}
\footnotetext[8]{ansible\#66201: \url{https://github.com/ansible/ansible/pull/66201}}
\footnotetext[9]{ansible\#66203: \url{https://github.com/ansible/ansible/pull/66203}}
\footnotetext[10]{home-assistant\#31783: \url{https://github.com/home-assistant/home-assistant/pull/31783}}
\textbf{Approach:}
We create a pull request and submit the patch to the project developers.
Listing~\ref{sr:oss_example} shows an example of the generated patches.
After generating the patches we write the pull request message manually to follow each project pull request template.
\begin{figure}[t]
    \begin{lstlisting}[caption=Examples of the merged source code changes in a real OSS project, label=sr:oss_example,language=diff]
// A merged source code changes for TensorFlow (C++)
// https://github.com/tensorflow/tensorflow/pull/35600
- runner_ = [](std::function<void()> fn) { fn(); };
+ runner_ = [](const std::function<void()>& fn) { fn(); };

// A merged source code changes for ansible (C++)
// https://github.com/ansible/ansible/pull/66201
- name: myTestNameTag
+ Name: myTestNameTag
\end{lstlisting}
\end{figure}

To filter unused patterns as in RQ3, we define the \textit{Pattern Frequency} from number of the pattern matched files.
We define the $Consequent Files$ as the project files that include consequent patterns.
Also, we define the $Condition Files$ as the project files that include condition patterns.
We suggest the highest $Pattern Frequency$ pattern to the target projects, and make a patch by using the measure of pattern frequency.
In this paper, we used patterns that have more than a 0.50 \textit{Pattern Frequency}.
In other words, the more than half of the project files followed our target patterns.
\begin{equation}
P(Frequency)=\frac{|Consequent Files \setminus Condition Files|}{|Consequent Files \cup Condition Files|}
\end{equation}	

\begin{figure}
    \centering
    \includegraphics[width=\linewidth]{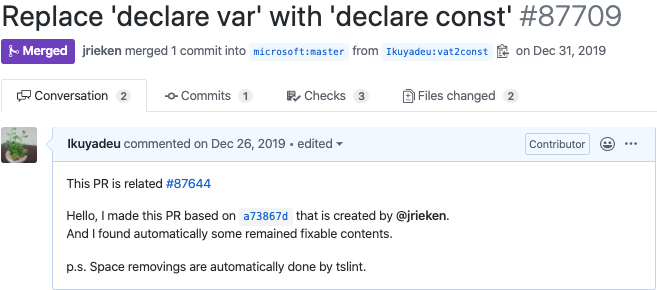}
    \caption{Submitted pull request on Microsoft/vscode project by using the~\devreplay~pattern information}
    \label{fig:pullmessage}
\end{figure}

To exclude dependency on individual skills for pull request submissions and to reduce the reviewers' efforts, we put the original change information from the matched patterns on pull requests.
Table~\ref{tab:pulls_info} shows the commit information from the patterns and the original pull request information from the original ``Commit ID'' information.
Using this information, Figure~\ref{fig:pullmessage} shows an example of the pull request message on the ``Microsoft/vscode'' project that used the information ``Commit Message'' (\texttt{Replace 'declare var...' with 'declare const'})  ``Pull Request ID'' (\texttt{This PR is related \#87644}), ``Commit ID'' (\texttt{this PR is based on a73867d}), ``Author GitHub ID''(\texttt{created by @jreiken}).

\begin{table*}[t]
\centering
\scriptsize
\caption{Patterns information and examples that are used on Microsoft/vscode\#87709}\label{tab:pulls_info}
\begin{tabular}{lll}
 Commit elements & Purpose & Example \\\hline
 Commit author & Identify the original commit author & Johannes Rieken \\
 Diff & Understand the change contents & \texttt{- declare var Buffer: any;}\\
&& \texttt{+ declare const Buffer: any;} \\
 Committed date & Filter the old patterns & 2019/12/25 12:15 \\
 Commit ID & Identify the original commit & a73867d \\
 File name & Identify the target language & buffer.ts \\\\
 Pull request & & \\
 elements & Purpose & Example \\\hline
 Repository name& Identify the target project & Microsoft/vscode \\
 Pull request status &  & Merged / Closed / Open \\
 Pull request ID & Refer to the original pull request & \#87644 \\
 Author GitHub ID & Refer to the original author & @jrieken  \\
 Reviewer GitHub ID & Refer to the original reviewer & @sandy081 \\
 Commit Message & Describe change reasons & replace 'declare var...' with\\&& 'declare const...', related \\&& \#87644 \\
\end{tabular}
\end{table*}

\textbf{Result:}

8 of 10 pull requests have already merged.
All of the 8 pull requests have been reviewed by the same developer with the original commit author or reviewer.
In the state-of-the-art machine learning, APR tools do not have the information of the original commit and author~\cite{Tool_phoenix}.
We believe notifying the original reviewer of our tool makes it is easy for reviewer recommendation.

Most project members did not send the change request to the accepted pull requests except for ``Thank you'' and ``lgtm'' (Looks good to me).
Only the home-assistant project member suggested the source code fix to adjust the code format and test suites.
9 our generated patch changed only one or two lines, and that did not have an impact for the test and some issues.

However, some projects did not accept the pull requests that where not related to the source code behavior.
for a rejected pull request comment in Visual Studio Code project, a project member said, ``\textit{The original change makes sure not to register a listener, this PR is purely cosmetic and those changes, esp when touching multiple feature area owners, aren't accepted}''.
Our tool does not take into account whether this change will fix the any issues or not.

\fbox{
\begin{tabular}{l}RQ4 Answer: \\
Project members merged 8 of 10 automatically\\
generated pull requests. Most pull requests\\
were reviewed by the same author and refereed\\
by \devreplay.
\end{tabular}
}
\section{Related Works}\label{sec:related_work}

Several research areas are related to this study.
In this section, we introduce existing static analysis tools, automatic program repair tools, benchmark data sets, change distilling techniques.

\subsection{Coding conventions}

To help developers use a common implementation style, some large software projects provide their own coding guideline.
The coding guideline usually includes general conventions for programming languages such as PEP8~\cite{van2001pep}, CERT C, and MISRA C.
These conventions and styles focus on the programming language. Unlike existing conventions, we extract hidden coding conventions from the git history.

\subsection{Static analysis tools}~\label{subsec:sat}


We developed a static analysis tool that suggests code changes based on the target project change history. In this section, we introduce existing static analysis tools that focus on language, large projects, and major API specific coding conventions.

Programming language-specific static analysis tools are  used to detect common implementation convention issues.
\textit{Checkstyle} is used for detecting whether or not java source code follows its coding conventions~\cite{smit2011code}.
In addition, some convention tools such as \textit{Pylint}~\cite{thenaultpylint}, \textit{Flake8}~\cite{flake8} and \textit{pydocstyle}~\cite{pydocstyle} check the format of python coding convention violations. Also, some violations can be fixed automatically.

in specific cases, converters of Python2 to Python3 are published~\cite{malloy_ESEM_2017}.
Even through Python3 was released in 2008, the converters are still updating in 2020s~\url{https://github.com/python/cpython/tree/master/Lib/lib2to3}.
\devreplay~can easily solve function replacing and change pattern sharing before tool updating.

\subsection{Automatic Program Repair (APR) tools}~\label{subsec:apr}

An atomic change operator program repair tools modifies a program in a single place of its Abstract Syntax Tree (AST) to passing the test suites.
The template-based program repair tools modify the program by more than 100,000 change history data such as in git.

Test suites-based APR tools use test suites to fix the buggy code~\cite{Tool_angelix, Tool_cvc4, Tool_enum, Tool_semfix, Tool_nopol}.
 These tools distinguished between declaration and body part changes, and define atomic change types based on an AST (Abstract Syntax tree).
 Since ASTs are rooted trees and these source code entities are either sub-ASTs or leafs, the basis for source code changes are elementary tree edit operations.
These techniques may include patterns that are irrelevant to bug fixes.

To extract useful and reliable patterns focusing on fix changes template-based APR tools have been published.~\cite{Tool_par, Xuan_SANER2016, Long_FSE2017, Tool_elixir,Tool_APIREC,Tool_sydit}.
Template-based APR tools fix source code based on code change patterns.
Recent research collected specific bugs fix patterns such as static analysis violations~\cite{Tool_avatar, Tool_avatar_pre, Tool_phoenix}.
Static analysis tools are used to fix source code on the coding process, and static analysis tools are used on the editor~\cite{Tool_findbugs, Tool_clang}.

Recent automated program repair techniques detect source code fix patterns.
To evaluate these patterns, many researchers have submitted more than 10 pull requests to real open source projects.
These approaches require work by open source developers, and knowledge from extracted patterns cannot be shared with among developers.
After submitting the pull request, our tools can help developers share useful pattern with real open source projects in the future.
As an example, \textsc{PAR} generates source code fix templates from the 60K+ human-written Java changes~\cite{Tool_par}.
However, ethere are only two ffective templates; namely, \textit{Null Pointer Checker} and \textit{Expression Adder/Remover/Replacer}. These fixes can be avoided by just using modern program languages, static analysis tools, or IDEs~\cite{Monperrus_ICSE2014}.

Most APR tools only focused on one major programming languages such as Java or C.
We apply~\devreplay~to Java and C test failures, and we suggest the source code changes also in TypeScript and Python projects that cannot detect the issues before execution.

\subsection{Repair benchmark}
In this study we used IntroClass and CodeFlaws that are C repair benchmarks.
As a major Java benchmark, Defects4j~\cite{Data_defects4j} is used by several APR tools.
Also, RepairThemAll~\cite{RepairThemAll} supports benchmark comparisons for Java-based APR tools.
Our tool can suggest code changes to the Java files.
For a review data set, Tufano et al. collected the code change data set from Gerrit Code Review~\cite{Data_CodeChangeML}.
However, we did not use that benchmark due to the target bugs related to multiple lines and files that are out of scope when using static analysis tools.

\subsection{Change distilling}

To improve static analysis tools and APR tools, change distilling techniques are published~\cite{Allamanis_TSE2018,Negara_ICSE2014,Nguyen_ICSE2019,Fluri_TSE2007,Barbez_ICSME2019,Hanam_FSE2016,Zhang_ICSE2019,Tool_Revisar,Tool_fixminer}.
FixMiner~\cite{Tool_fixminer} and SysEdMiner~\cite{Tool_SysEdMiner} have similar functions as our tool.
They detect code change patterns from git repositories that fixed with static code analysis tool violations.
One of the large differences between~\devreplay~and other tools is that we distill code changes, even if the target source code does not have any violation.s
Unlike existing tools, \devreplay~refers fix patterns that can be edited and shared within multi-develop environment. 

Change distilling techniques are believed to improve the programming language and static analysis tools.
BugAID~\cite{Hanam_FSE2016} distilled frequently appeared JavaScript bug patterns from 105K commits in 134 projects.
As in the previous study, most bug fix patterns that occurred were de-referenced non-values.
According to manual check, these major bug patterns can be detectable by using a static-type system~\cite{JS2TS}.
Unlike these famous bug fixes, our goal is to distill project-specific fix patterns and suggest source code changes with the change evidence.

We designed \devreplay~as the three independent processes with pattern generation, pattern matching and pattern suggesting.
By following the \devreplay~pattern format, existing change distilling result can be applied as part of pattern generation part, and can be used on the supported editor and code review.

\subsection{Code Review}

\devreplay~helps in the code review process.
In the code review process, patch authors and reviewers often discuss and propose solutions with each other to revise patches~\cite{Tsay_ISFSE2014}.
Reviewers spend much time verifying the proposed code changes through code review manually~\cite{Rigby_ICSE2011, Bosu_ESEM2014}.
Reviewers detect not only large impact issues, but some code reviews are improve code based on coding conventions~\cite{Tao_ICSME2014,Boogerd2008,smit2011code}.
75\% of discussions for revising a patch are about software maintenance and 15\% are about functional issues\cite{Beller_MSR2014,Czerwonka_ICSE2015}.
\devreplay~helps to introduce code reviewers' conventions to novice patch authors.

\section{Threats to Validity and Limitation} \label{sec:threats_to_validity}

\subsection{Internal Validity}

Only adding lines' changes is within the scope of our paper.
In the pattern file, adding change is difficult in understanding the matching condition.

We did not focus on the null check that has been confirmed as one of the useful change patterns.
Our tool focuses on the patterns that are appeared in the change history, and our target data set did not have null check issues.

In the evaluation of RQ1, we compared~\devreplay~with existing tools that have completely different use cases.
\devreplay~is automatically recommended in the code editor while existing tools actively execute bug fix commands.
Therefore, the actual work time of the developer differs between the two tools.
In this analysis, we only compared accuracy of the each tool recommendations.

In the wild evaluations for RQ3 and RQ4, we do not have any baseline.
Similar studies exist, but the projects they target have included many problems before \devreplay~is used.
For example, in the existing studies, the source code was modified using a traditional static analysis tool and test cases.
However, 5 of 8 of our targeted projects already use the systems that automatically detect static analysis tools and test warnings and propose modifications.
We focus the 
We have proposed fixes for projects that are already well managed with static analysis tools and testing.

We collected the code change data from 1 day/week/month commit histories. We used this threshold based on an existing study method~\cite{Nguyen_ICSE2019}. This existing study and our evaluation focus on a large software project that has dozens of commits daily. In the implementation, this time period can be adjusted for the project size.

\subsection{External Validity}

\devreplay~only targets the one hunk changes. In addition, our tool will work during code editing.
A large data set such as Defects4j~\cite{Data_defects4j} are outside of the scope of this paper.
Unlike existing Java program repair studies,~\devreplay~will not check the test suite and code behavior.
Also, multiple files changes are out of our scope too. These changes are difficult to covered by regular expressions.

\section{Summary}\label{sec:summary}

In this study, we presented a static analysis tool, \devreplay, that suggests the source code changes based on a projects' git history.
As a result, our tool outperformed the state-of-the-art APR tools. Also, 80\% of change suggestions are accepted by language-cross open source software projects.
For future work, we will improve the code change pattern generating method to filter out the unused or duplicated patterns.

\section{Attachment}\label{sec:attachment}
\begin{figure}[ht]
    \centering
    \includegraphics[width=\linewidth]{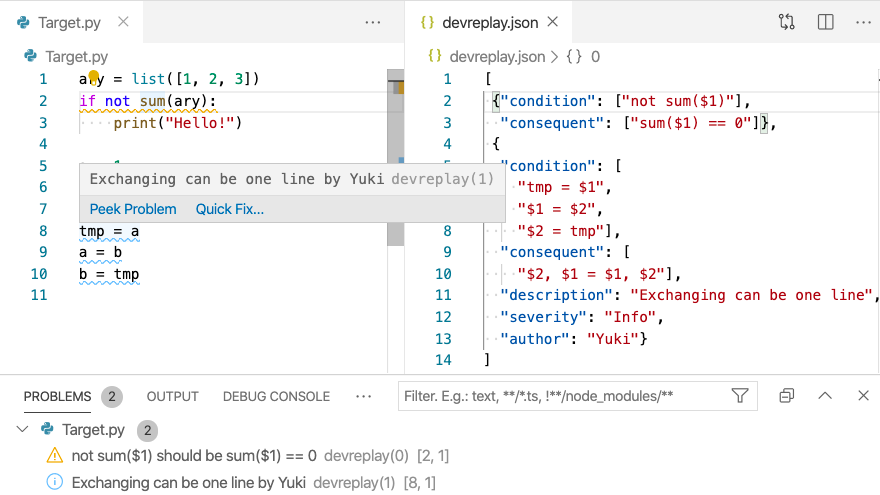}
    \caption{\devreplay~on the Visual Studio Code editor. The left side is the target python code. The right side is the pattern file. The bottom tab shows the warning messages}
    \label{fig:vscode_implementation}
\end{figure}
Developers can fix source code by using the~\devreplay~not only on the command-line interface, but also during the code editing process on the editor and code review process on GitHub.
Using these independent constructions, users and tool developers can improve \devreplay~by themselves.
As an example, if someone develops new change distilling method with an output format that follows the~\devreplay, the distilled patterns can be used on the same platform as the~\devreplay.

\subsection{Usage on source code editor}
Using the matched patterns, \devreplay~can suggest source code changes in several environments.
Below real tool use method examples are shown on the Visual Studio Code editor.
A \devreplay~user can fix source code using the following process
\begin{enumerate}
\item User installs the~\devreplay~plugin to the Visual Studio Code at~\url{https://marketplace.visualstudio.com/items?itemName=Ikuyadeu.devreplay}.
\item User generates a patterns file as ``devreplay.json'', then puts the file on the project root path.
\item User opens the project source code files by the Visual Studio Code. Figure~\ref{fig:vscode_implementation} shows the target source code on the left side. Also, right side shows the a simple ``devreplay.json'' which has the two patterns. Finally, the bottom tab shows the warning messages.
\item \devreplay~shows the warnings bya  colored wavy line based on ``devreplay.json'' patterns. On the left side of Figure~\ref{fig:vscode_implementation}, \devreplay~shows the warning messages on the line 2 and lines 8-10. In the second pattern, the warnings color and description can change by editing the ``devreplay.json''. ``severity'' element changes with the wavy line colors. ``description'' and ``author'' elements change the warning description.
\item Users clicks the warning lines light bulb symbol and push ``Fix by DevReplay'' and \devreplay~fixes the target lines. 
\end{enumerate}

\begin{figure*}[t]

\begin{subfigure}{0.5\textwidth}
\includegraphics[width=\linewidth]{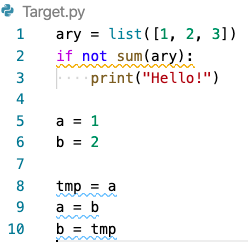} 
\caption{Pre-fixed}
\label{fig:prechanged}
\end{subfigure}
\begin{subfigure}{0.5\textwidth}
\includegraphics[width=\linewidth]{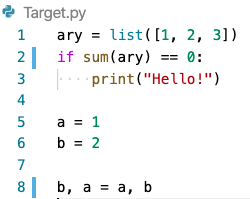}
\caption{Fixed}
\label{fig:changed}
\end{subfigure}

\caption{Fixed contents by~\devreplay~on the Visual Studio Code editor}
\label{fig:vscode_changed}
\end{figure*}
\begin{figure*}[t]

\begin{subfigure}{0.5\textwidth}
\includegraphics[width=\linewidth]{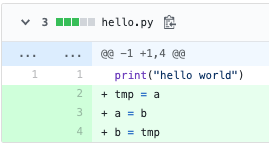} 
\caption{Submitted pull request}
\label{fig:submitted_pull}
\end{subfigure}
\begin{subfigure}{0.5\textwidth}
\includegraphics[width=\linewidth]{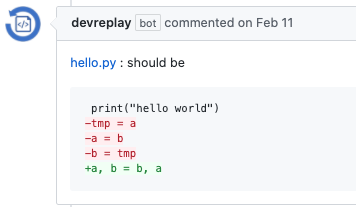}
\caption{\devreplay suggestion}
\label{fig:generated_review}
\end{subfigure}
\caption{Reviewed contents by~\devreplay~on the GitHub code review}
\label{fig:auto_codereview}
\end{figure*}
Figure~\ref{fig:vscode_changed} shows the changed contents. Target file fixed two points, First pattern means the clearing the \textit{if}-statement expression, second one convert the value swap process to the one line.

This function are not only available at the Visual Studio Code, it also can be used on the other editors such as Vim by executing language server protocol implementation at~\url{https://www.npmjs.com/package/devreplay-server}.

\subsection{Usage on code review}

Also, the project team can maintain source code consistency by using the implementation of code review at~\url{https://github.com/marketplace/dev-replay}.

Project team automatically review source code by the following process

\begin{enumerate}
\item Project team installs the \devreplay~bot from the GitHub market place.
\item Project team generates patterns as ``devreplay.json'', and then put these patterns file on the project root path.
\item Some developers submit a pull request to the project.
\item \devreplay~automatically suggests the source code through the pull request.
\item The user fixes the source code to avoid the \devreplay~fix without a real project members' review.
\end{enumerate}


\bibliographystyle{IEEEtran}      
\bibliography{bibfile}

\end{document}